\documentstyle[psfig,amssymb]{mn}
\title[Red AGN in XMM-Newton/SDSS fields]{Red Active Galactic
Nuclei in XMM-Newton/SDSS fields}     
 
\author[Georgakakis, Georgantopoulos \& Akylas]
  {A. E. Georgakakis$^{1,2}$\thanks{age@imperial.ac.uk}, I. Georgantopoulos$^1$ \& A. Akylas$^1$
  \\ \\
  $^1$Institute of Astronomy \& Astrophysics, National Observatory of
  Athens, I. Metaxa \& V. Pavlou, Athens, 15236, Greece \\ 
  $^2$Imperial College of Science Technology and Medicine, Blackett
  Laboratory, Prince Consort Rd, 2BZ SW7, London, UK\\ 
}
\begin{document}
\maketitle  

\begin{abstract}
In this paper we combine archival and proprietary  {\it XMM-Newton}
observations (about 5\,deg$^{2}$) that overlap with the Sloan Digital
Sky Survey to explore the nature of the moderate-$z$ X-ray population.
We focus on X-ray sources with optically red colours ($g-r>0.4$),
which we argue are important for understanding the origin of the
X-ray background. Firstly, these systems constitute a significant
fraction, about 2/3, of the $z\la1$ X-ray population  to the limit $f(
\rm 2-8\,keV ) \approx 2 \times 10^{-14} \rm erg \, s^{-1} \,cm^{-2}$.
Secondly, their luminosity function under evolution of the form
$\propto (1+z)^{3}$ suggests that they could be responsible for about
17 per cent of the diffuse X-ray background  to $z=1$. Thirdly, their
stacked X-ray spectrum in the range 1-8\,keV is consistent with a
power-law distribution with $\Gamma  \approx 1.4$ (without fitting
intrinsic absorption), i.e. similar to the diffuse X-ray
background. We find that the optically red X-ray population comprises
a mixed bag of objects, both obscured ($N_H > 10^{22} \rm \, cm^{-2}$)
and unobscured ($ N_H<10^{22} \rm \, cm^{-2}$), with a wide range of
X-ray luminosities up $L_X\approx10^{44}\, \rm erg \, s^{-1}$. We
argue that dilution of the AGN light by the host galaxy  may play a
role in shaping the continuum optical emission of this
population. Finally, we explore a possible association of these
sources and the moderate-$z$ red ($J-Ks>2$\,mag) AGNs identified in
the Two Micron All Sky Survey (2MASS). The median $N_H$ of the red
X-ray sources studied here is $\approx \rm  10^{21} \, cm^{-2}$, lower
than that found for the 2MASS AGNs, suggesting different populations.    
\end{abstract}


\section{Introduction}\label{sec_intro}

The origin of the diffuse X-ray background (XRB) remains one of the
most debated issues of X-ray astronomy. Deep X-ray surveys with the
new generation X-ray missions, the {\it  Chandra} and the {\it
XMM-Newton} have revolutionised this field demonstrating that the bulk
of the XRB can be resolved into discrete point sources (e.g. Brandt et
al. 2001; Giaconni et  al. 2002). These surveys have confirmed
previous results that luminous  ($L_X \ga  10^{44} \rm \, erg \,
s^{-1}$) broad emission-line QSOs peaking at redshifts $z\approx1.5-2$
are  undoubtedly a major component of the XRB, especially at energies
below $\approx 2$\,keV (e.g. Lehmann et al. 2001). These sources
alone however, cannot produce the bulk of the XRB. For example their
steep X-ray spectra ($\Gamma \approx 1.9$) are inconsistent with  the
spectral shape of the X-ray background ($\Gamma \approx 1.4$;
e.g. Gendreau et  al. 1995).  An additional population of X-ray
sources is clearly required to account for the XRB properties.  

There is in particular, accumulating evidence suggesting that a large
fraction of the XRB, particularly at energies $\rm >2\,keV$ may arise
at redshifts $z \la 1$ in moderate luminosity systems ($L_X \la 
10^{44} \rm \, erg \, s^{-1}$). For example, previous missions at
hard energies ($>2$\,keV;  {\it HEAO1 A-2}, {\it Ginga}), although
they lacked imaging capabilities, suggest a statistically significant
cross-correlation signal between the detected XRB fluctuations and
nearby galaxy catalogues (e.g. ESO, UGC, IRAS). This implies that
about 30 per cent of the XRB could be produced at low-$z$ (Jahoda  et
al. 1991; Lahav et al. 1993; Miyaji et al. 1994; Carrera  et
al. 1995). In the {\it Chandra} and the {\it XMM-Newton} era, a
non-negligible fraction of the hard X-ray population has been
identified with optically extended sources suggesting $z\la 1$
galaxies (Koekemoer et al. 2002; Grogin et al. 2003;  Georgantopoulos
et al. 2004). This is further confirmed by 
spectroscopic follow-up observations, which show a redshift
distribution with  a peak at $z \la 1$ for hard X-ray selected samples
(Barger et al. 2002; Rosati et al. 2002; Fiore et al. 2003;
Georgantopoulos et al. 2004; Georgakakis et al. 2004a).    
Although the difficulty to spectroscopically identify optically faint
sources ($R>25$\,mag) may skew the above distribution to lower $z$
(e.g.  Treister et al. 2004), it is nevertheless accepted that these
optically faint systems (about 25 per cent in the {\it  Chandra} deep
fields) cannot drastically modify the observed distribution.

The evidence above suggests that parallel to the deep X-ray surveys
targeting the high-$z$ Universe, it is important that we also study
the nature of the moderate-$z$ X-ray population. This may indeed hold
important clues for the origin of the XRB and for interpreting deeper
X-ray samples.  

A large fraction of the $z \la 1$ hard X-ray sources
in particular, are associated with optically red galaxies suggesting
either dust reddening (e.g. Seyfert-2s) or continuum emission
dominated by stars rather than the central AGN (Georgantopoulos et
al. 2004; Georgakakis et al. 2004a). There is also
evidence that this population has an average X-ray spectrum similar to
that of the X-ray background ($\Gamma \approx 1.4$), further
underlying its significance for XRB studies (Georgantopoulos  et
al. 2004).  

In this paper we explore the nature of the optically red nearby
($z<1$) X-ray sources and their significance in shaping the XRB by
combining public {\it XMM-Newton} data with the Sloan Digital Sky
Survey (SDSS; York et al. 2000) to exploit the uniform 5-band optical
photometry and spectroscopy available in this area. Wide field
coverage is essential to probe large enough volume at moderate
redshifts to provide a representative sample of this class of
sources. The {\it XMM-Newton} with 4 times the {\it Chandra}
field-of-view provides an ideal platform for such a study. Throughout
this paper we adopt $\rm H_{o} = 70 \, km \, s^{-1} \, Mpc^{-1}$, $\rm
\Omega_{M} = 0.3$ and $\rm \Omega_{\Lambda} = 0.7$. 

\section{The X-ray data}\label{sec_survey}
In this paper we use 28 {\it XMM-Newton} archival observations
selected to overlap with the second data release of the SDSS (DR2; 
Stoughton et al. 2002) and with a proprietary period that expired
before September 2003. A total of 8 of these fields are part of the
{\it XMM-Newton}/2dF survey (Georgakakis et al. 2003, 2004b;
Georgantopoulos et al. 2004), while the remaining 20 pointings are
presented by Georgantopoulos \& Georgakakis (2005). The exposure times
are in the range 2-67\,ks with a median of about 15\,ks. The Galactic
column density in the direction of these fields varies between $\rm
1.3 - 13 \times 10^{20} \, cm^{-2}$ with a median of about $\rm 2 \times 
10^{20} \, cm^{-2}$. A full description of the data reduction, the
event file generation and the X-ray image production is
presented by Georgantopoulos \& Georgakakis (2005).     

\begin{figure}
\centerline{\psfig{figure=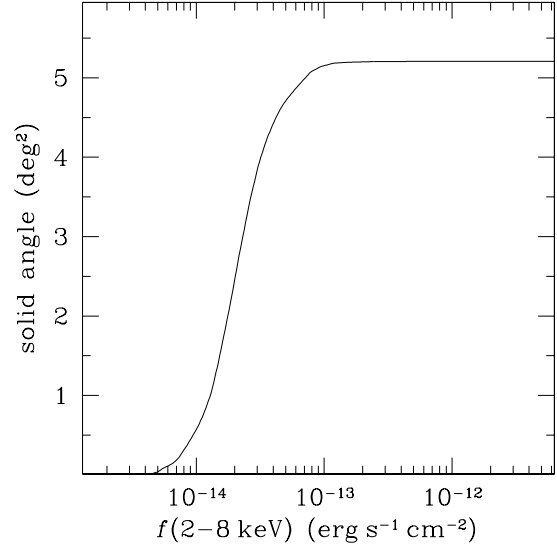,width=3in,angle=0}}
\caption
 {Solid angle as a function of 2-8\,keV flux ($6\sigma$) for our
 survey.  
 }\label{fig_area_curve}
\end{figure}

For this paper the source extraction is performed in the 2-8\,keV
merged  PN+MOS images (when available) using the {\sc ewavelet} task
of {\sc sas} with a detection threshold of $6\sigma$. This choice of
threshold is to minimise spurious detections in the final
catalogue. The extracted sources for each field  were visually
inspected and spurious detections clearly associated with CCD gaps,
hot pixels or lying close to the edge of the field of view were 
removed. We further exclude from the final catalogue the target of a
given {\it XMM-Newton} pointing (e.g. nearby galaxies or clusters) and
a total of 7 sources that appear extended on the {\it XMM-Newton}
EPIC images and are clearly  associated with diffuse cluster
emission. Fluxes are estimated using an 18\,arcsec radius aperture
corresponding to an encircled energy fraction of about 70 per cent at 
1.5\,keV. For the spectral energy distribution we adopt a power-law
with  $\Gamma=1.8$ and Galactic column appropriate for each
field. Using $\Gamma=1.4$ instead of $\Gamma=1.8$, typical of the mean
spectrum of the XRB and the red AGN studied here (see section 4.2),
has a minimal effect on the estimated fluxes (about  8 per cent) and
does not affect the results presented here. For  the background
estimation we use the background maps generated as a by-product of the
{\sc ewavelet} task of {\sc sas}. The final hard X-ray selected sample
comprises a total of 507 sources above the $6\sigma$ flux  limit $f_X
(\rm 2 - 8 \, keV) = 5 \times 10^{-15} \, erg \, s^{-1} \,
cm^{-2}$. The area curve giving the cumulative area of our survey as a
function of limiting flux is shown in Figure \ref{fig_area_curve}.

\section{The sample}
\subsection{Selection criteria}

We use the SDSS DR2 catalogue to optically identify the hard X-ray
selected sources following the method of Downes et al. (1986), as
described in Georgakakis et al. (2004b) to calculate the probability,
$P$,  a given candidate is a spurious identification. Here we apply an 
upper limit in the search radius, $r<7\rm\,arcsec$ and a cutoff in the
probability, $P<0.05$, to limit the optical  identifications to those
candidates that are least likely to be spurious alignments.

In this study we use  $6\,\sigma$ hard X-ray selected
sources with  2-8\,keV flux $f_X(\rm 2  - 8 \, keV) > 2 \times
10^{-14} \, erg \, s^{-1} \, cm^{-2}$. This is to ensure sufficient
photon-statistics to perform X-ray spectral analysis for most sources
and to minimise the number of objects without optical
identifications. Out of 230 sources above the $6\,\sigma$ detection
threshold and  $f_X(\rm 2 - 8 \, keV) > 2 \times 10^{-14} \, erg \,
s^{-1} \, cm^{-2}$ a total of 189 have optical counterparts to the
SDSS limit (82 per cent completeness). The choice of the X-ray flux
limit above is a trade-off between maximum optical identification
completeness and sufficiently large sample size to avoid small number
statistics.  

\begin{figure}
\centerline{\psfig{figure=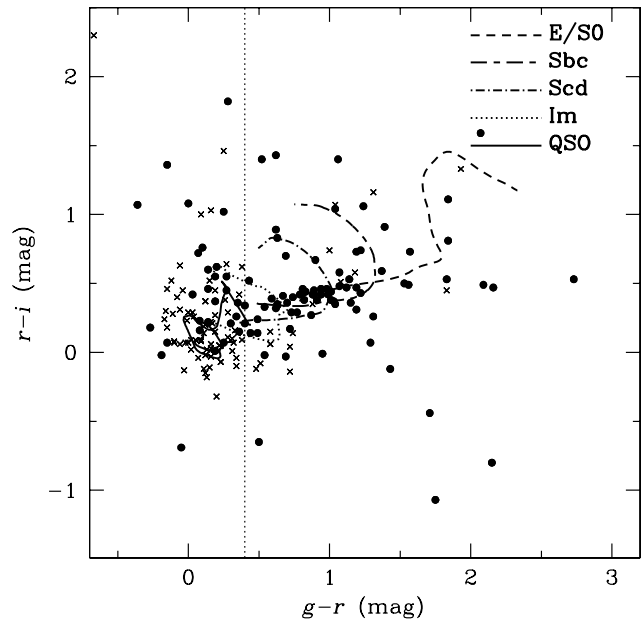,width=3.5in,angle=0}}
\caption
 {$r-i$ against $g-r$ colour for optically identified $6\sigma$
 sources with $f_X(\rm 2 - 8 \, keV) > 2 \times 10^{-14} \, erg \,
 s^{-1} \, cm^{-2}$. Filled circles and crosses are for optically
 extended and point-like X-ray sources  respectively. The expected
 colours of 
 different galaxy types (E/S0, Sbc, Scd, Im) using the template SEDs
 from Coleman, Wu \& Weedman (1980) are also shown for redshifts
 $z=0-1$. Also plotted is a QSO template SED for $z=0-3$ (Cristiani et
 al. 2004). Our colour selection $g-r>0.4$ is 
 shown with the vertical dotted line.  
 }\label{fig_cc}
\end{figure}

From the sample above we select sources with red optical colours,
$g-r>0.4$\,mag, to exclude bluer broad-line QSOs. Optically red sources
have been identified in  non-negligible numbers in recent surveys
(e.g. Koekemoer et al. 2002) and
may play a key role in shaping the diffuse  X-ray background. Indeed,
this population may be associated with obscured AGNs and/or the
 population of X-ray Bright Optically Normal Galaxies
(XBONGs; Comastri et al. 2002) that are also suggested to harbor
deeply buried AGNs (but see Georgantopoulos \& Georgakakis 2005). 
Figure \ref{fig_cc} plots the optical colours of the hard X-ray
selected sample with the expected tracks for different galaxy types
(E/S0, Sbc, Scd, Im; Coleman, Wu \& Weedman 1980) and optically
selected QSOs (Cristiani et al. 2004). In this
figure it is clear that the colour cut $g-r>0.4$\,mag  selects
against typical broad line QSOs providing a sample that is dominated
by sources with galaxy-like colours. Nevertheless, high-$z$ QSO
($z>2$; e.g. Richards et al. 2002; Kitsionas et al. 2005) or reddened
QSOs (e.g. Wilkes et al. 2002; White et al. 2003) with SEDs different
to those of optically selected ones may still exist within our
sample. Such sources are expected to be identified by their unresolved
(e.g. point-like) optical light profile. We also caution the reader
that the colour cut $g-r=0.4$ also  eliminates from  the sample low-
($z\la0.2$) or high-$z$ ($z\ga0.8$) systems with SEDs similar to
irregular (Im) type galaxies.

A total of 102 hard X-ray selected sources fulfil the criteria
$f_X(\rm 2 - 8 \, keV) > 2 \times 10^{-14} \, erg \, s^{-1} \,
cm^{-2}$ and $g-r>0.4$. Figure \ref{fig_cc} shows that most of them
(total of 84) have extended optical light profile suggesting
moderate-$z$  ($\la1$) systems where the central AGN does not
dominate the optical light profile. The 18 sources with unresolved
optical light profile are associated with either high-$z$ QSOs or
Galactic stars. We note however, that the SDSS star-galaxy separation
is reliable at the 95\% confidence limit to $r=21$\,mag  and becomes
less robust at fainter magnitudes. A total of 25 out of 102 sources in
the sample are fainter that this magnitude limit. Also there are only
29 [$\approx 25$ per cent, 29/(29+84)] optically extended sources with 
colours bluer than  $g-r=0.4$ within the $f_X(\rm 2 - 8 \, keV) > 2
\times 10^{-14} \, erg \, s^{-1} \, cm^{-2}$ subsample. The red
subpopulation, comprising 75 per cent of the optically extended
X-ray sources, clearly dominates at moderate-$z$.

\subsection{Redshift estimation}

Spectroscopic redshifts are available for 24 hard X-ray selected
sources with $g-r>0.4$ from either the SDSS or our own follow-up
campaign of the {\it XMM-Newton}/2dF survey. A total of 22 of the 24
sources are optically extended and are assigned moderate redshifts
($z<1$). The remaining two X-ray sources are optically unresolved and
are identified with a Galactic star and a $z=1.335$ broad-line QSO 
(Georgantopoulos et al. 2004).  For the spectroscopically unidentified
sources in the sample we use photometric methods to estimate
redshifts.    

In particular for systems with extended  optical light profile we use
the SDSS-DR2 photometric redshifts that are based on galaxy templates
(Csabai et al. 2002). As a byproduct of the photometric redshift
method each source is assigned a best-fit SED which is a continuous
parameter between 0 and 1. The two extreme values 0 and 1  correspond
to ellipticals and actively starforming (irregulars)  systems
respectively. Galaxies with intermediate best fit SEDs are assigned
photometric types between 0 and 1. 

\begin{figure} 
\centerline{\psfig{figure=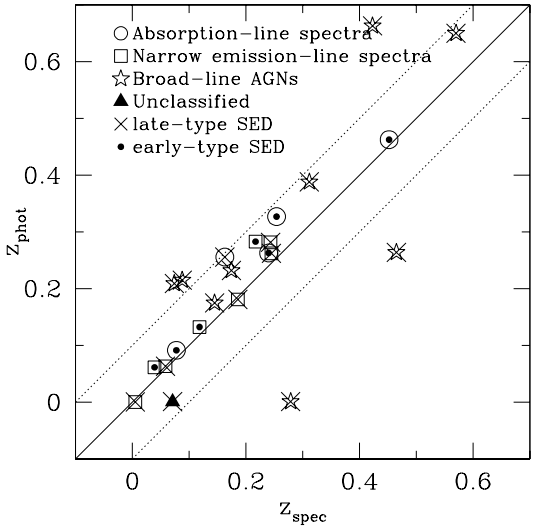,width=3in,height=3in,angle=0}} 
\caption
 {Photometric against spectroscopic redshift estimates for the
 hard X-ray selected sources with available spectroscopic
 observations. Open circles are systems with absorption line
 spectra, open squares correspond to sources with narrow-emission line 
 spectra and stars are broad line AGNs. The source in our sample (\#84
 in Table 1) without spectral classification is shown with the filled
 triangle. A cross on top of a symbol indicates a late-type best fit SED
 (photometric type $>$0.2) to the photometric data, while small  dots
 are for early type SEDs (photometric type $<$0.2).The dashed lines
 are the $\delta z= \pm 1$ envelope around the $z_{spec}=z_{phot}$
 continuous line. 
 }\label{fig_zz}    
\end{figure}

Previous studies on photometric redshifts of X-ray sources suggest
that galaxy templates work well for systems with red optical colours,
similar to those studied here (Barger et al. 2002, 2003; Mobasher et
al. 2004; Georgakakis et al. 2004a; Kitsionas et
al. 2005). The success rate using galaxy templates for these sources is
demonstrated in Figure \ref{fig_zz} which compares the photometric and
spectroscopic redshift estimates for the optically extended objects
with available spectroscopic observations.  The agreement is good with
$| (z_{phot} - z_{spec}) | / (1+z_{spec}) \approx 0.06$. A number of
sources significantly deviate ($\delta z>0.1$) from the
$z_{spec}=z_{zphot}$ relation in Figure \ref{fig_zz}. All of them show
broad emission-lines and it is likely that AGN light is
contributing to their optical continuum. Figure \ref{fig_zz} also
shows there is a fair agreement between the observed spectral
classification (emission, absorption line) and the  galaxy type of the
best fit SED (early, late types).   For absorption-line
objects 4 out of 5 are best-fit by early-type SEDs  ($<0.2$). For
narrow emission-line galaxies and broad-line systems 5/7 and 9/9 are
best-fit by late type SEDs respectively. One spectroscopically
identified source has no spectral classification in   Figure
\ref{fig_zz} (see below for details).

\begin{figure}
\centerline{\psfig{figure=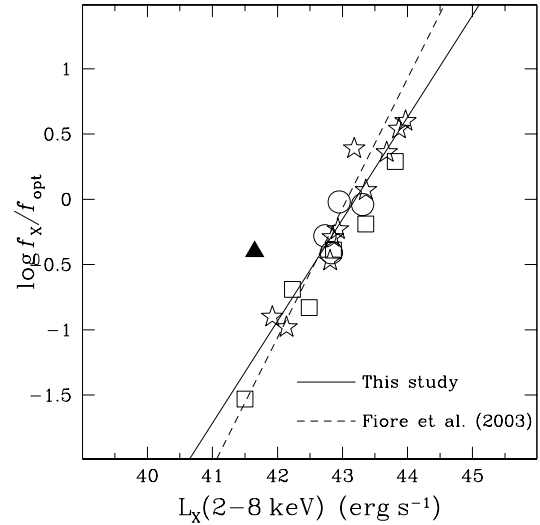,width=3in,angle=0}}
\caption
 {2--8\,keV X-ray luminosity against X-ray--to--optical flux ratio for
 red X-ray sources with available optical spectroscopy. Open circles
 are systems with absorption line spectra, open squares correspond to
 sources with narrow-emission line spectra and stars are broad line
 AGNs. The source in our sample (\#84 in Table 1) without spectral
 classification is 
 shown with the filled triangle. The continuous line is the best-fit
 relation to the data. The dashed line is the best-fit relation of
 Fiore et al. (2003) for their sample of non type-I AGNs.
 }\label{fig_fxfo_lx}
\end{figure}

A small number of 4 optically faint ($r\approx22$\,mag; see Table
\ref{tab_sample1}) galaxies in our sample are assigned unrealistically
low photometric redshifts, $z\approx0.001$, by the SDSS algorithm. For
these systems we use the relation between X-ray luminosity and
X-ray--to--optical flux ratio (see next section for the definition of
$\log f_X/f_{opt}$) suggested by Fiore et al. (2003) to get a rough
redshift estimate. Figure \ref{fig_fxfo_lx} plots this relation for
the optically red sources in our sample with spectroscopic redshift
available. Using this relation we estimate $z\approx1-2$ for the 4
optically faint sources. These systems are marked in Table
\ref{tab_sample1}.  The redshift uncertainty using the $1\sigma$ rms
scatter around the best-fit relation in Figure \ref{fig_fxfo_lx} is
estimated to be $\delta z / (1+z) \approx 0.2$. 

\begin{figure} 
\centerline{\psfig{figure=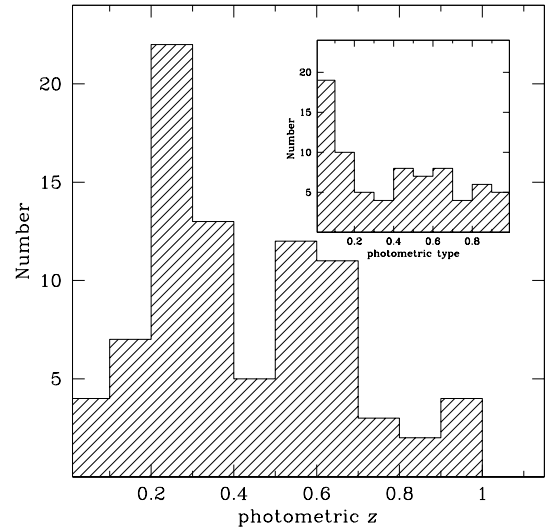,width=3in,height=3in,angle=0}} 
\caption
 {Redshift (photometric or spectroscopic) and photometric type
 distribution for the red optically extended X-ray
 sources. 
 }\label{fig_distribution}    
\end{figure}

Figure \ref{fig_distribution} plots the redshift and spectral type
distribution of the optically extended subsample. About half of the
sources lie at $z \la 0.4$ and have best fit spectral types $<0.4$.    

A total of 18 X-ray sources with red colours have point-like optical
light profile and are most likely associated with either Galactic
stars, or high-$z$ QSOs. Using the photometric methods described by
Kitsionas et al. (2005) we find that 5 of these sources are best fit
by stellar templates with the remaining assigned photometric redshifts
$z>1.3$. These 18 sources will not be considered in the rest of the
paper. 

We also exclude from the analysis source \#47 in Table 1 at $z=0.005$ with
$L_X \approx 2 \times 10^{39}\, \rm erg \, s^{-1}$. This system has
X-ray--to--optical flux ratio $<-2$ (see Figure \ref{fig_fxfo} below)
and has been classified `normal' galaxy by  Georgantopoulos et
al. (2005) with X-ray emission associated with stellar processes. For
completeness this is included in Table 1 (see below). The final sample
used in the analysis comprises a total of 83 systems with
$g-r>0.4$ and extended  optical light profile. For luminosity
estimates photometric redshifts are used unless spectroscopic
redshifts are available 

\subsection{X-ray spectral analysis}
We explore the X-ray spectral properties of our sample using  the {\sc
xspec} v11.2 package. For sources
with small number of net counts we use the C-statistic technique
(Cash 1979) specifically developed to extract information from low
signal-to-noise ratio spectra. The data are grouped to have at least
one count per bin. We adopt an absorbed  power-law model ({\sc
wabs*pow}) to constrain the absorbing column density $N_H$ by
fixing  the power law index  to $\Gamma=1.8$, i.e. in-between radio
loud and radio quiet AGNs.  For sources with sufficient net counts
(about 150) we perform standard $\chi^{2}$ spectral
fitting. The data are  grouped to have a minimum of 15 counts per bin
to ensure that Gaussian statistics  apply and we require that the
spectrum has at least 10  spectral bins. In the case of $\chi^{2}$
analysis an absorbed power-law ({\sc wabs*pow}) is fit to the data
yielding both the $N_H$ and the power-law photon index
$\Gamma$. For both the $\chi^{2}$ and the C-statistic
analysis the fit was performed in the 0.2-8\,keV energy range where
the sensitivity of the {\it  XMM-Newton} is the highest. The estimated
errors correspond to the 90 per cent confidence level. 

The column densities estimated above are in the observer's frame and
therefore lower than the rest-frame $N_H$ because the $k$-effect
shifts the absorption turnover to lower energies. The relation between
intrinsic and observed column density scales approximately as
$(1+z)^{2.65}$ (e.g. Barger et al. 2002). This redshift correction is
applied to all sources, after subtracting from the observed
$N_H$ the appropriate Galactic column in the direction of the
source. The rest-frame $N_H$ is then used to estimate the unabsorbed
X-ray luminosities.  

The optical and X-ray properties of the hard X-ray selected sample
with $g-r>0.4$ and $f_X(\rm 2 - 8 \, keV) > 2 \times 10^{-14} \, erg
\, s^{-1} \, cm^{-2}$ used in this paper is presented in Table
\ref{tab_sample1} which has the following format:   

{\bf 1.} Identification number.  

{\bf 2-3.} Right ascension and declination of the X-ray centroid
position in J2000. 

{\bf 4.} SDSS $r$-band magnitude of the optical counterpart. 


{\bf 5.} Probability, $P$, the optical counterpart is a chance
coincidence.

{\bf 6.} 2-8\,keV X-ray flux in $\rm erg\, s^{-1} \, cm^{-2}$.

{\bf 7.} X-ray--to--optical flux ratio, $\log f_X / f_{opt}$, estimated
from the relation  
\begin{equation}\label{eq2}
\log\frac{f_X}{f_{opt}} = \log f_X(2-8\,{\rm keV}) +
0.4\,r +5.60.
\end{equation}
The equation above is derived from the X-ray--to--optical flux
ratio definition of Stocke et al. (1991) that involved 0.3-3.5\,keV
flux and $V$-band magnitude. These quantities are converted to
2-8\,keV flux and $r$-band magnitude adopting a mean colour
$V-R=0.7$, the transformations of Fukugita et al. (1996) and a
power-law X-ray spectral energy distribution with index
$\Gamma=1.8$. 

{\bf 8.} Spectroscopic redshift measurement if available.

{\bf 9.} Spectroscopic classification. AB is for absorption line
systems, NL is for narrow emission-line sources and BL signifies broad
emission-line optical spectra. Source \#84 has a spectroscopic
redshift estimate from the NED but no spectral classification is
available. 

{\bf 10.} Photometric redshift estimate. 

{\bf 11.} 2-8\,keV X-ray luminosity in units of $\rm erg\, s^{-1}$,
corrected for X-ray absorption. 

{\bf 12.} Rest-frame column density, $N_H$, estimated from
the spectral fitting analysis in units of $\rm 10^{22}\, cm^{-2}$.

{\bf 13.} Power-law spectral index, $\Gamma$, for those sources with
sufficient net counts (about 150) to perform $\chi^2$ analysis. The
median $\Gamma$ for these sources is $\approx 1.8$.

\begin{table*}
\scriptsize
\begin{center}
\begin{tabular}{l ccc cc ccc ccc c}
\hline
ID &
$\alpha_X$ &
$\delta_X$ &
$r$ &
$P$ &
$f_X(\rm 2-8\,keV)$ &
$\log f_X / f_{opt}$ &
$z_{spec}$    &
spectral    &
$z_{phot}$  &
$\log L_X$ &
$N_H$ &
$\Gamma$\\

  &
(J2000) &
(J2000) &
(mag) &
(\%) &
($\rm 10^{-14}\,cgs$) &
       & 	
       &
class  &  
       &
(2--8\,keV) &
($\rm 10^{22} \, cm^{-2}$) &
 \\
\hline
01$^{}$ & 00 43 50.23 & $+$00 57 50.01 & 20.90 &  0.13 & $14.80\pm1.47$ & $+$1.12 & $ - $ & $-$ & 0.126 & 42.80 &  $0.17_{-0.08}^{+0.08}$ & $2.13_{-0.25}^{+0.30}$\\
02$^{}$ & 01 52 04.83 & $+$01 08 02.43 & 22.88 &  4.41 & $4.53\pm1.05$ & $+$1.40 & $ - $ & $-$ & 0.600 & 43.79 &  $0.42_{-0.29}^{+0.40}$ & $-$\\
03$^{}$ & 01 52 30.38 & $+$00 57 02.17 & 22.37 &  4.90 & $3.49\pm0.65$ & $+$1.08 & $ - $ & $-$ & 0.354 & 43.14 & $0.12_{-0.12}^{+0.11}$ & $-$\\
04$^{}$ & 01 52 38.35 & $+$01 09 24.23 & 19.11 &  0.52 & $4.82\pm0.92$ & $-$0.07 & $ - $ & $-$ & 0.286 & 43.08 &  $1.69_{-0.50}^{+2.57}$ & $-$\\
05$^{}$ & 01 53 07.35 & $+$01 05 53.79 & 19.35 &  0.19 & $6.16\pm1.00$ & $+$0.12 & $ - $ & $-$ & 0.262 & 43.10 &  $ < 0.03$ & $-$\\
06$^{}$ & 01 59 57.60 & $+$00 33 10.85 & 18.78 &  0.02 & $31.00\pm4.01$ & $+$0.60 & $0.312$ & BL & 0.388 & 43.97  & $ < 0.03$ & $2.10_{-0.15}^{+0.25}$\\
07$^{}$ & 02 00 25.21 & $+$00 29 17.18 & 19.07 &  0.43 & $5.84\pm2.01$ & $+$0.00 & $ - $ & $-$ & 0.434 & 43.57 &  $ < 0.09$ & $-$\\
08$^{}$ & 02 00 28.97 & $+$00 28 48.03 & 17.87 &  0.10 & $10.40\pm2.58$ & $-$0.23 & $0.174$ & BL & 0.232 & 42.93  & $ < 0.01$ & $-$\\
09$^{}$ & 02 40 49.24 & $-$08 09 42.72 & 20.17 &  0.01 & $2.21\pm0.50$ & $+$0.01 & $ - $ & $-$ & 0.563 & 43.41 &  $ < 0.03$ & $-$\\
10$^{}$ & 02 41 34.68 & $-$08 09 59.72 & 21.17 &  0.49 & $2.58\pm0.59$ & $+$0.47 & $ - $ & $-$ & 0.239 & 42.63 &  $0.08_{-0.08}^{+0.19}$ & $-$\\
11$^{}$ & 02 56 27.34 & $+$00 07 36.81 & 20.95 &  1.29 & $3.91\pm0.84$ & $+$0.57 & $ - $ & $-$ & 0.345 & 43.17 &  $1.08_{-0.80}^{+0.78}$ & $-$\\
12$^{}$ & 02 56 45.33 & $+$00 00 32.27 & 19.25 &  0.07 & $5.53\pm1.02$ & $+$0.04 & $ - $ & $-$ & 0.392 & 43.45 &  $ < 0.04$ & $-$\\
13$^{}$ & 03 02 06.60 & $-$00 00 04.88 & 21.16 &  0.64 & $2.44\pm0.37$ & $+$0.45 & $ - $ & $-$ & 0.342 & 42.96 &  $0.32_{-0.15}^{+0.25}$ & $-$\\
14$^{1}$ & 03 02 34.75 & $+$00 01 07.51 & 21.48 &  0.47 & $4.26\pm0.34$ & $+$0.82 & $ - $ & $-$ & 1.36 & 44.24 &  $ < 0.06$ & $1.83_{-0.12}^{+0.07}$\\
15$^{}$ & 03 03 15.04 & $-$00 01 03.58 & 20.86 &  0.45 & $2.26\pm0.42$ & $+$0.29 & $ - $ & $-$ & 0.532 & 43.37 &  $7.73_{-2.98}^{+4.81}$ & $-$\\
16$^{}$ & 03 03 20.28 & $+$00 06 05.76 & 18.89 &  0.02 & $4.28\pm0.45$ & $-$0.21 & $ - $ & $-$ & 0.248 & 42.89 &  $2.60_{-1.01}^{+-0.04}$ & $1.80_{-0.89}^{+0.78}$\\
17$^{1}$ & 03 37 52.40 & $+$00 22 11.09 & 21.03 &  1.15 & $2.68\pm0.90$ & $+$0.43 & $ - $ & $-$ & 1.04 & 43.75 &  $ < 0.04$ & $-$\\
18$^{}$ & 03 38 37.61 & $+$00 20 47.14 & 21.59 &  1.76 & $4.86\pm1.17$ & $+$0.92 & $ - $ & $-$ & 0.577 & 43.78 &  $4.13_{-2.96}^{+5.07}$ & $-$\\
19$^{}$ & 03 39 01.02 & $+$00 19 50.76 & 20.04 &  0.23 & $2.34\pm0.85$ & $-$0.01 & $ - $ & $-$ & 0.484 & 43.28 &  $8.75_{-5.85}^{+39.14}$ & $-$\\
20$^{}$ & 08 30 26.83 & $+$52 46 01.94 & 20.14 &  0.08 & $5.57\pm0.39$ & $+$0.40 & $ - $ & $-$ & 0.661 & 43.98 &  $ < 0.15$ & $1.79_{-0.14}^{+0.25}$\\
21$^{}$ & 08 31 08.61 & $+$52 38 39.57 & 21.70 &  0.39 & $4.03\pm0.30$ & $+$0.88 & $ - $ & $-$ & 0.772 & 44.00 &  $1.71_{-0.84}^{+0.55}$ & $1.89_{-0.40}^{+0.15}$\\
22$^{}$ & 08 31 17.50 & $+$52 48 56.00 & 22.81 &  0.63 & $3.50\pm0.23$ & $+$1.26 & $ - $ & $-$ & 0.604 & 43.68 &  $ < 0.03$ & $1.63_{-0.06}^{+0.09}$\\
23$^{}$ & 08 31 39.11 & $+$52 42 06.87 & 15.70 &  0.02 & $3.88\pm0.28$ & $-$1.53 & $0.059$ & NL & 0.063 & 41.50 & $22.86_{-3.58}^{+4.66}$ & $-$\\
24$^{}$ & 09 16 45.13 & $+$51 41 44.93 & 17.97 &  0.27 & $11.00\pm1.30$ & $-$0.17 & $ - $ & $-$ & 0.305 & 43.49 & $ < 0.03$ & $1.85_{-0.16}^{+0.14}$\\
25$^{1}$ & 09 17 44.88 & $+$51 37 39.54 & 22.40 &  1.51 & $2.01\pm0.48$ & $+$0.86 & $ - $ & $-$ & 1.97 & 44.29 & $0.93_{-0.86}^{+1.09}$ & $-$\\
26$^{}$ & 09 18 04.00 & $+$51 41 16.02 & 17.67 &  0.25 & $3.16\pm-0.93$ & $-$0.83 & $0.186$ & NL & 0.181 & 42.49  & $5.88_{-1.18}^{+13.34}$ & $-$\\
27$^{}$ & 09 18 58.69 & $+$51 43 42.38 & 20.03 &  0.74 & $2.37\pm0.67$ & $-$0.01 & $ - $ & $-$ & 0.526 & 43.38 &  $ < 0.03$ & $-$\\
28$^{}$ & 09 19 17.17 & $+$51 41 27.71 & 21.91 &  2.69 & $3.97\pm0.96$ & $+$0.96 & $ - $ & $-$ & 0.802 & 44.03 &  $1.50_{-0.93}^{+1.45}$ & $-$\\
29$^{}$ & 09 33 05.68 & $+$55 06 01.96 & 21.25 &  0.62 & $2.41\pm0.50$ & $+$0.48 & $ - $ & $-$ & 0.332 & 42.92 &  $2.86_{-1.22}^{+2.11}$ & $-$\\
30$^{}$ & 09 33 24.32 & $+$55 18 30.65 & 22.29 &  0.57 & $2.25\pm0.44$ & $+$0.86 & $ - $ & $-$ & 0.745 & 43.70 &  $2.77_{-1.05}^{+0.87}$ & $-$\\
31$^{}$ & 09 33 32.58 & $+$55 04 59.85 & 21.82 &  1.26 & $2.16\pm0.49$ & $+$0.66 & $ - $ & $-$ & 0.119 & 41.90 &  $1.37_{-0.61}^{+1.15}$ & $-$\\
32$^{}$ & 09 33 48.10 & $+$55 18 46.56 & 20.62 &  0.02 & $14.49\pm0.67$ & $+$1.00 & $ - $ & $-$ & 0.523 & 44.15 & $0.22_{-0.07}^{+0.11}$ & $1.63_{-0.09}^{+0.09}$\\
33$^{}$ & 09 33 52.67 & $+$55 26 15.98 & 20.12 &  0.61 & $2.27\pm0.50$ & $+$0.00 & $ - $ & $-$ & 0.549 & 43.40 &  $ < 0.12$ & $-$\\
34$^{}$ & 09 33 59.84 & $+$55 10 01.03 & 19.43 &  $<0.01$ & $2.38\pm0.43$ & $-$0.25 & $ - $ & $-$ & 0.411 & 43.13  & $18.08_{-5.41}^{+5.92}$ & $-$\\
35$^{}$ & 09 34 58.59 & $+$61 12 33.50 & 17.71 &  0.11 & $13.30\pm0.93$ & $-$0.19 & $0.245$ & NL & 0.262 & 43.36 &  $1.53_{-0.46}^{+0.45}$ & $-$\\
36$^{1}$ & 09 35 01.09 & $+$55 23 52.10 & 21.73 &  0.83 & $2.00\pm0.69$ & $+$0.59 & $ - $ & $-$ & 1.42 & 43.94 &   $< 0.06$ & $-$\\
37$^{}$ & 09 35 34.01 & $+$61 23 52.47 & 19.05 &  0.17 & $2.68\pm0.35$ & $-$0.34 & $ - $ & $-$ & 0.264 & 42.74 &  $5.96_{-1.55}^{+2.25}$ & $-$\\
38$^{}$ & 09 35 35.54 & $+$61 19 21.38 & 19.13 &  0.35 & $4.95\pm0.42$ & $-$0.05 & $ - $ & $-$ & 0.287 & 43.09 &  $0.02_{-0.02}^{+0.04}$ & $2.48_{-0.12}^{+0.16}$\\
39$^{}$ & 09 36 08.61 & $+$61 30 28.73 & 19.91 &  0.05 & $3.16\pm0.52$ & $+$0.06 & $ - $ & $-$ & 0.246 & 42.75 &  $0.09_{-0.09}^{+0.15}$ & $2.06_{-0.44}^{+0.37}$\\
40$^{}$ & 09 36 53.29 & $+$61 26 27.27 & 20.81 &  1.67 & $2.26\pm0.70$ & $+$0.27 & $ - $ & $-$ & 0.591 & 43.47 &  $ < 0.07$ & $-$\\
41$^{}$ & 12 30 58.77 & $+$64 17 27.23 & 21.20 &  0.80 & $2.31\pm0.32$ & $+$0.44 & $ - $ & $-$ & 0.631 & 43.55  & $0.62_{-0.18}^{+0.41}$ & $-$\\
42$^{}$ & 12 33 25.99 & $+$64 14 52.26 & 19.54 &  1.10 &
$2.81\pm0.61$ & $-$0.13 & $ - $ & $-$ & 0.294 & 42.87 &  $ < 0.10$ &
$2.11_{-0.28}^{+0.43}$\\
\hline
\end{tabular}
\end{center}
\caption{
Optical and X-ray  properties of the hard X-ray selected sample.
}\label{tab_sample1}
\normalsize
\end{table*}

\begin{table*}
\contcaption{}
\scriptsize
\begin{center}
\begin{tabular}{l ccc cc ccc ccc c}
\hline

ID &
$\alpha_X$ &
$\delta_X$ &
$r$ &
$P$ &
$f_X(\rm 2-8\,keV)$ &
$\log f_X / f_{opt}$ &
$z_{spec}$    &
spectral&
$z_{phot}$  &
$\log L_X$ &
$N_H$ &
$\Gamma$\\

  &
(J2000) &
(J2000) &
(mag) &
(\%) &
($\rm 10^{-14}\,cgs$) &
       & 
       &
class   & 
       &
(2-8\,keV) &
($\rm 10^{22} \, cm^{-2}$) &
  \\
\hline
43$^{}$ & 12 44 24.83 & $-$00 24 37.88 & 18.32 & 0.04 & $6.07\pm0.57$ & $-$0.28 & $0.179$ & AL & 0.267 & 42.73  & $ < 0.01$ & $1.72_{-0.07}^{+0.11}$\\
44$^{}$ & 12 44 29.30 & $-$00 34 16.51 & 21.16 & 1.07 & $4.50\pm0.42$ & $+$0.71 & $ - $ & $-$ & 0.956 & 44.27 &  $2.04_{-0.50}^{+0.96}$ & $-$\\
45$^{}$ & 12 44 57.61 & $-$00 16 17.07 & 17.10 & 0.01 & $3.80\pm0.49$ & $-$0.98 & $0.118$ & BL & 0.133 & 42.14 &  $ < 0.05$ & $1.74_{-0.15}^{+0.21}$\\
46$^{}$ & 12 45 05.74 & $-$00 31 43.48 & 19.30 & 1.51 & $2.16\pm0.22$ & $-$0.34 & $ - $ & $-$ & 0.105 & 41.77 &  $1.25_{-0.38}^{+0.69}$ & $-$\\
47$^{}$ & 12 45 32.14 & $-$00 32 05.01 & 13.12 & 0.01 & $2.09\pm0.27$ & $-$2.83 & $0.005$ & NL & 0.001 & 39.16 &  $0.05_{-0.05}^{+0.09}$ & $1.58_{-0.60}^{+0.23}$\\
48$^{}$ & 12 45 57.66 & $-$00 30 37.03 & 18.87 & 0.60 & $3.25\pm0.47$ & $-$0.34 & $ - $ & $-$ & 0.522 & 43.50 &  $0.12_{-0.12}^{+0.06}$ & $1.64_{-0.16}^{+0.23}$\\
49$^{}$ & 13 02 56.20 & $+$67 37 37.43 & 19.73 & 0.02 & $2.01\pm-0.63$ & $-$0.20 & $ - $ & $-$ & 0.221 & 42.46  & $11.47_{-5.84}^{+10.77}$ & $-$\\
50$^{}$ & 13 03 36.83 & $+$67 30 26.58 & 19.53 & 0.11 & $15.09\pm0.87$ & $+$0.58 & $ - $ & $-$ & 0.386 & 43.87  & $0.02_{-0.02}^{+0.02}$ & $2.29_{-0.05}^{+0.13}$\\
51$^{}$ & 13 03 57.60 & $+$67 28 32.24 & 18.39 & 0.04 & $3.72\pm0.68$ & $-$0.47 & $0.243$ & BL & 0.282 & 42.81  & $ < 0.02$ & $1.96_{-0.19}^{+0.29}$\\
52$^{}$ & 13 05 50.30 & $+$67 39 20.92 & 19.87 & 0.33 & $11.00\pm1.64$ & $+$0.58 & $ - $ & $-$ & 0.638 & 44.24  & $ < 0.29$ & $1.87_{-0.27}^{+0.37}$\\
53$^{}$ & 13 40 38.65 & $+$00 19 18.75 & 19.63 & 0.11 & $8.80\pm1.43$ & $+$0.39 & $0.244$ & BL & 0.549 & 43.18  & $ < 0.01$ & $-$\\
54$^{}$ & 13 40 45.17 & $-$00 24 01.72 & 20.19 & 4.03 & $9.48\pm1.61$ & $+$0.65 & $ - $ & $-$ & 0.215 & 43.09 &  $0.41_{-0.13}^{+0.38}$ & $-$\\
55$^{}$ & 13 41 18.05 & $-$00 23 20.79 & 19.64 & 0.47 & $12.20\pm1.30$ & $+$0.54 & $0.423$ & BL & 0.663 & 43.87  & $ < 0.24$ & $1.82_{-0.25}^{+0.43}$\\
56$^{}$ & 13 41 28.35 & $-$00 31 20.36 & 20.64 & 0.02 & $13.50\pm1.73$ & $+$0.98 & $ - $ & $-$ & 0.619 & 44.29 &  $5.57_{-1.63}^{+2.69}$ & $-$\\
57$^{}$ & 13 41 34.39 & $+$00 28 07.52 & 20.27 & 0.22 & $2.01\pm0.61$ & $+$0.01 & $ - $ & $-$ & 0.676 & 43.56 &  $ < 0.04$ & $-$\\
58$^{}$ & 13 41 40.46 & $+$00 15 43.88 & 18.60 & 0.80 & $3.55\pm0.97$ & $-$0.41 & $0.254$ & AL & 0.327 & 42.83  & $0.01_{-0.01}^{+0.16}$ & $-$\\
59$^{}$ & 13 42 12.05 & $+$00 29 49.45 & 20.42 & 0.60 & $3.99\pm1.33$ & $+$0.36 & $0.570$ & BL & 0.650 & 43.68 & $0.05_{-0.05}^{+0.82}$ & $-$\\
60$^{}$ & 13 43 47.54 & $+$00 20 21.45 & 18.14 & 0.37 & $12.40\pm2.44$ & $-$0.04 & $0.240$ & AL & 0.263 & 43.31  & $ < 0.08$ & $-$\\
61$^{}$ & 13 43 51.13 & $+$00 04 38.00 & 16.75 & 0.16 & $6.25\pm1.84$ & $-$0.90 & $0.074$ & BL & 0.209 & 41.92 &  $ < 0.01$ & $-$\\
62$^{}$ & 13 44 20.14 & $+$00 04 17.10 & 19.98 & 2.45 & $4.76\pm1.09$ & $+$0.26 & $ - $ & $-$ & 0.302 & 43.13 &  $0.65_{-0.32}^{+0.46}$ & $-$\\
63$^{}$ & 13 44 22.08 & $-$00 34 19.67 & 18.20 & 0.10 & $5.36\pm1.61$ & $-$0.39 & $0.217$ & NL & 0.283 & 42.86  & $0.09_{-0.09}^{+0.38}$ & $-$\\
64$^{}$ & 13 44 46.99 & $-$00 30 08.59 & 20.19 &  0.67 & $2.45\pm0.74$ & $+$0.06 & $ - $ & $-$ & 0.529 & 43.39 &  $3.16_{-2.34}^{+1.76}$ & $-$\\
65$^{}$ & 13 44 51.57 & $-$00 23 01.14 & 19.64 &  0.30 & $2.50\pm0.65$ & $-$0.14 & $ - $ & $-$ & 0.261 & 42.71 &  $1.42_{-0.89}^{+0.77}$ & $-$\\
66$^{}$ & 13 44 52.91 & $+$00 05 20.97 & 16.32 &  0.01 & $37.80\pm3.54$ & $-$0.29 & $0.088$ & BL & 0.215 & 42.85  & $ < 0.01$ & $2.39_{-0.11}^{+0.10}$\\
67$^{}$ & 13 44 58.03 & $-$00 36 00.23 & 19.39 & 1.36 & $8.69\pm1.64$ & $+$0.29 & $0.465$ & NL & 0.264 & 43.81  & $26.68_{-14.66}^{+30.60}$ & $-$\\
68$^{}$ & 13 44 58.46 & $+$00 16 22.52 & 18.02 &  0.26 & $3.13\pm1.21$ & $-$0.69 & $0.145$ & NL & 0.175 & 42.23  & $ < 0.01$ & $-$\\
69$^{}$ & 13 45 07.38 & $+$00 04 09.09 & 19.53 &  0.67 & $5.66\pm1.56$ & $+$0.16 & $ - $ & $-$ & 0.553 & 43.80  & $ < 0.03$ & $-$\\
70$^{}$ & 13 45 08.44 & $+$00 22 26.59 & 21.05 &  1.19 & $2.52\pm1.06$ & $+$0.42 & $ - $ & $-$ & 0.415 & 43.16  & $ < 0.02$ & $-$\\
71$^{}$ & 13 45 09.78 & $+$00 20 52.17 & 19.54 &  0.82 & $15.70\pm2.32$ & $+$0.61 & $ - $ & NL & 0.307 & 43.66  & $0.20_{-0.08}^{+0.33}$ & $-$\\
72$^{2}$ & 13 45 10.38 & $+$00 18 50.80 & 19.17 &  0.19 & $5.14\pm1.43$ & $-$0.02 & $0.243$ & AL & 0.294 & 42.95  & $ < 0.05$ & $-$\\
73$^{}$ & 13 48 59.53 & $+$60 15 01.25 & 19.25 &  0.75 & $6.05\pm0.58$ & $+$0.08 & $ - $ & $-$ & 0.404 & 43.51 &  $ < 0.02$ & $1.97_{-0.10}^{+0.19}$\\
74$^{}$ & 13 49 28.10 & $+$60 06 17.63 & 20.05 &  0.73 & $4.42\pm0.53$ & $+$0.26 & $ - $ & $-$ & 0.167 & 42.53 &  $ < 0.01$ & $1.88_{-0.13}^{+0.29}$\\
75$^{}$ & 15 43 02.46 & $+$54 09 12.98 & 21.55 & 2.51 & $9.13\pm1.23$ & $+$1.17 & $ - $ & $-$ & 0.614 & 44.12 &  $0.47_{-0.20}^{+0.20}$ & $-$\\
76$^{}$ & 15 44 23.51 & $+$54 04 16.20 & 20.66 &  0.43 & $6.93\pm0.90$ & $+$0.70 & $ - $ & $-$ & 0.604 & 43.98 &  $ < 0.39$ & $1.66_{-0.23}^{+0.46}$\\
77$^{}$ & 15 44 24.34 & $+$53 55 46.91 & 18.62 &  0.13 & $9.65\pm0.71$ & $+$0.03 & $ - $ & $-$ & 0.393 & 43.69 &  $6.92_{-1.09}^{+1.43}$ & $-$\\
78$^{}$ & 15 44 35.47 & $+$54 05 41.13 & 20.61 &  1.02 & $3.39\pm0.59$ & $+$0.37 & $ - $ & $-$ & 0.840 & 44.01 &  $ < 0.11$ & $-$\\
79$^{}$ & 17 00 47.13 & $+$64 23 03.81 & 20.59 &  0.01 & $4.53\pm4.16$ & $+$0.49 & $ - $ & $-$ & 0.624 & 43.83 &  $9.73_{-6.06}^{+17.63}$ & $-$\\
80$^{}$ & 23 37 14.41 & $+$00 14 15.06 & 19.32 &  0.27 & $2.61\pm0.74$ & $-$0.25 & $ - $ & $-$ & 0.269 & 42.75 &  $4.96_{-2.66}^{+5.17}$ & $-$\\
81$^{}$ & 23 37 31.77 & $+$00 25 59.87 & 19.07 &  $<0.01$ & $3.90\pm0.86$ & $-$0.17 & $ - $ & $-$ & 0.348 & 43.18 & $ < 0.02$ & $-$\\
82$^{}$ & 23 37 38.00 & $+$00 09 58.74 & 20.52 &  0.27 & $3.48\pm0.95$ & $+$0.34 & $ - $ & $-$ & 0.997 & 44.20 &  $2.09_{-0.99}^{+0.94}$ & $-$\\
83$^{}$ & 23 38 11.49 & $+$00 20 45.43 & 18.70 & 0.02 & $9.91\pm1.17$ & $+$0.07 & $0.279$ & BL & 0.001 & 43.36 &  $ < 0.07$ & $1.80_{-0.19}^{+0.30}$\\
84$^{3}$ & 23 53 52.04 & $-$10 15 26.64 & 18.61 &  0.68 & $3.58\pm0.75$ & $-$0.40 & $0.071$ & $-$ & 0.001 & 41.65 &  $0.27_{-0.12}^{+0.15}$ & $-$\\
\hline
\multicolumn{13}{l}{$^1$photometric redshift from the
$L_X- \log f_X/f_{opt}$ relation} \\
\multicolumn{13}{l}{$^2$source \#72 is classified Galactic star by
2QZ. Inspection of the optical spectrum however, suggests an early
type galaxy at $z=0.243$}\\
\multicolumn{13}{l}{$^3$source \#84 has no spectral classification in NED}\\
\end{tabular}
\end{center}
\normalsize
\end{table*}

\begin{figure}
\centerline{\psfig{figure=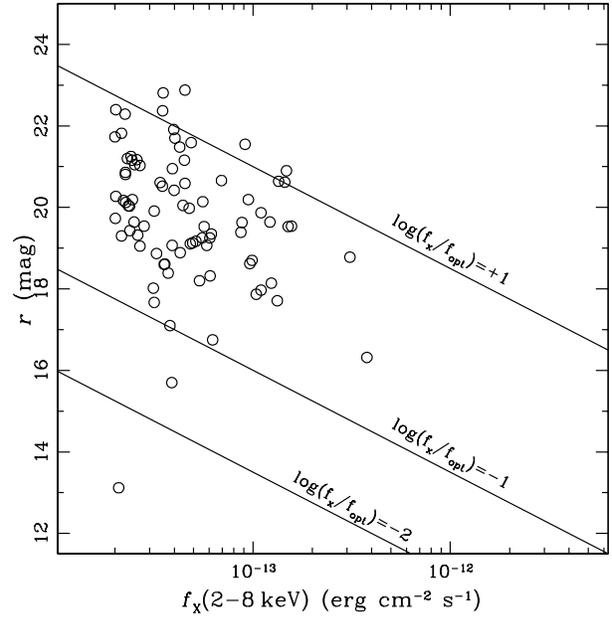,width=3.5in,angle=0}}
\caption
 {$r$-band magnitude against 2-8\,keV flux for the optically extended
 $g-r>0.4$ sources. The lines $\log f_X/f_{opt}=\pm1$
 delineate the region of the parameter space occupied by powerful
 unobscured AGNs.
 }\label{fig_fxfo}
\end{figure}

\begin{figure}
\centerline{\psfig{figure=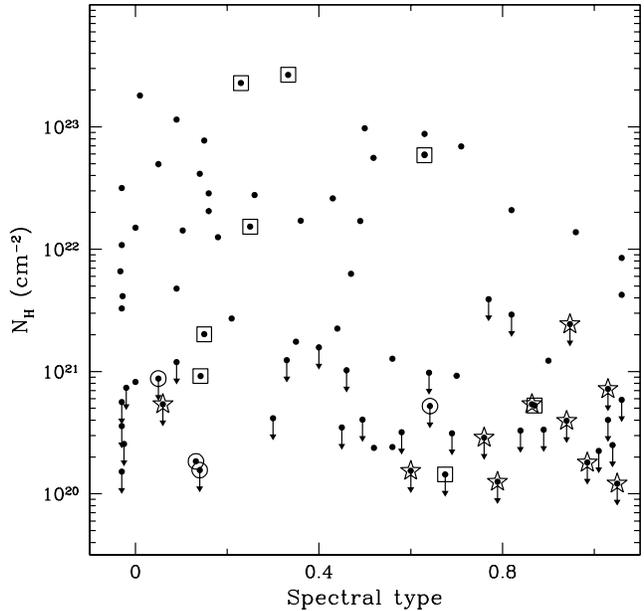,width=3.5in,angle=0}}
\caption
 {$N_H$  versus photometric redshift spectral type. Elliptical
 SEDs correspond to spectral type=0 while irregulars are assigned 
 type=1.  Upper limits in the hydrogen column density $N_H$ are
 shown with an arrow. A cross on top of a symbol is for sources with
 broad-line optical spectra. Similarly, open squares and open circles
 on top of a dot correspond to sources with narrow emission-line and
 absorption optical spectra respectively.
 }\label{fig_nh_stype}
\end{figure}

\begin{figure}
\centerline{\psfig{figure=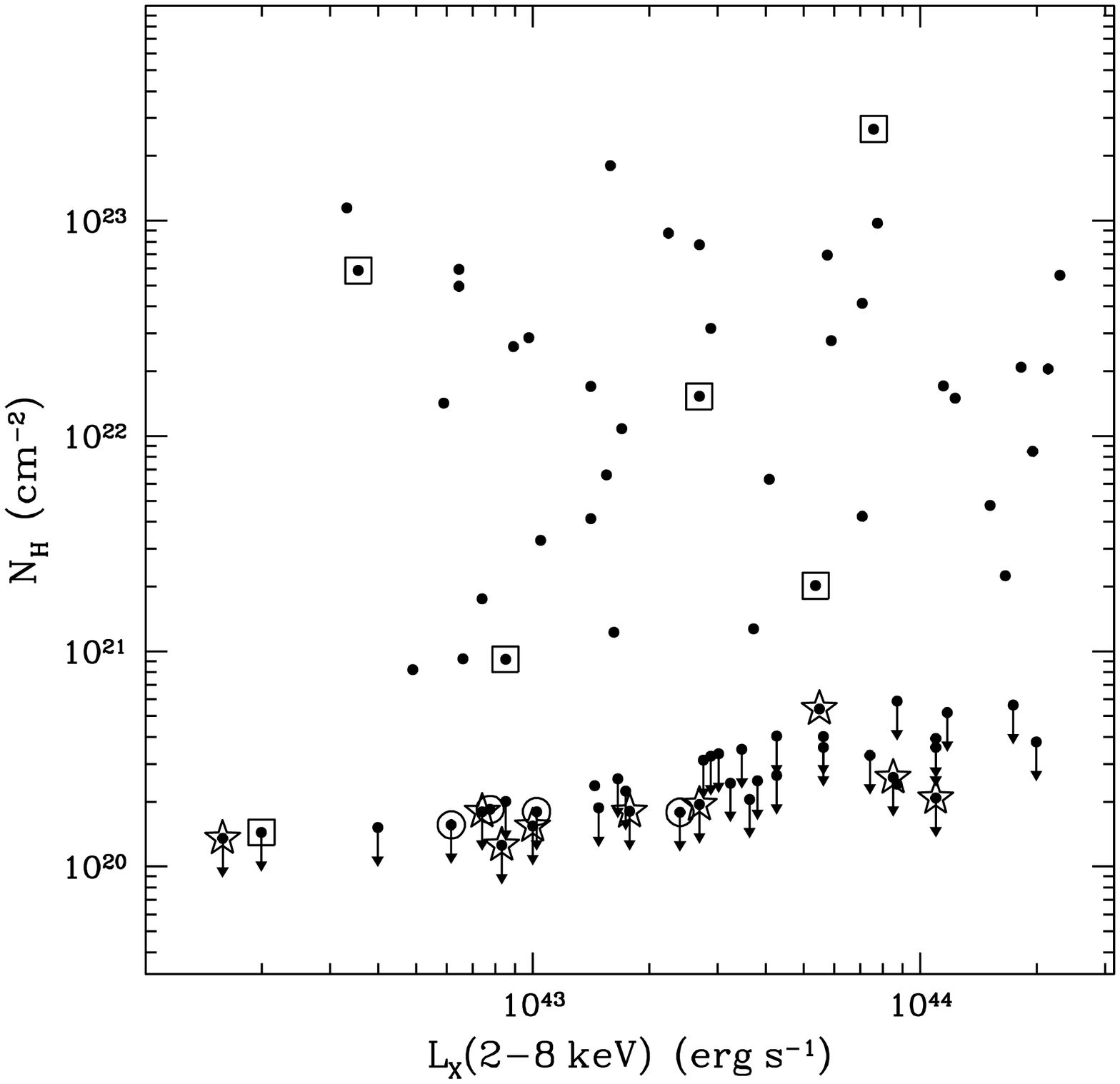,width=3.5in,angle=0}}
\caption
 {$N_H$  versus unabsorbed 2-8\,keV X-ray luminosity. The symbols are
 the same as in Figure \ref{fig_nh_stype}.
 }\label{fig_nh_lx}
\end{figure}

\section{Results}

\subsection{Optical and X-ray properties}
Figure \ref{fig_fxfo} plots $r$-band magnitude against 2-8\,keV  
X-ray flux. The  $\log (f_X/f_{opt})=\pm1$ lines in this figure
delineate the region of the parameter space occupied by AGNs (Stocke
et al. 1991). In this figure the red hard X-ray selected sources have
a wide range of X-ray--to--optical flux ratios. This may suggest
diverse AGN accretion rates, black hole masses, line-of-sight
obscurations and host galaxy properties.  

The present hard X-ray selected sample is indeed, heterogeneous in its
optical and X-ray properties. Figure \ref{fig_nh_stype} plots
rest-frame $N_H$ as a function of best-fit SED estimated as a
by-product of the photometric redshift estimation. The X-ray sources
studied in this paper are hosted by a wide range of galaxy types with
no strong correlation with intrinsic X-ray absorption. 
Figure
\ref{fig_nh_lx} demonstrates  that the unobscured 2-8\,keV X-ray
luminosities, providing a proxy to accretion rate and black hole mass,
take a wide range of values and are uncorrelated with the estimated
$N_H$.      

We further explore the nature of the red optically extended hard X-ray
selected sources using the optical spectroscopy available for a small  
but representative subsample (21 out of 83 sources). Most of the
narrow emission-line galaxies are associated with obscured systems
($N_H \ga 10^{21} \rm \, cm^{-2}$; see Figures \ref{fig_nh_stype},
\ref{fig_nh_lx}), suggesting Seyfert-2 type activity. It is likely
that the  optical emission of these systems is dominated by the host
galaxy rather than the central engine (Barger et et al. 2002;
Georgakakis et al. 2004a). One of the narrow 
emission-line galaxies nevertheless, does not show evidence for X-ray
obscuration ($N_H\la \rm 5 \times 10^{20} \, cm^{-2}$) although the
X-ray luminosity of $L_X (\rm 2 - 8 \,keV) \approx 2\times10^{42} \,
erg \, s^{-1}$ is higher than the maximum that starburst activity is
believed to be able to produce ($\rm \approx 10^{42} \, erg \,
s^{-1}$; Zezas, Georgantopoulos \& Ward 1998; Moran et al. 1999). A
similar class of sources are the `composites' discussed by Moran et
al. (2002) and Georgantopoulos, Zezas \&  Ward (2003) where
star-formation  is suggested to outshine the AGN. Although  this is a
possible scenario we cannot use diagnostic emission line ratios for
classification because the $\rm H\beta$ line is diluted by
instrumental features.    

Sources with absorption-line optical spectra have $L_X(\rm 2-8 \,keV)
> 5\times10^{42} \rm \, erg \, s^{-1}$ and are likely to belong to the class
of X-ray Bright Optically Normal Galaxies (XBONGs; Comastri et
al. 2002; Georgantopoulos \&  Georgakakis 2005). None of  these
sources show evidence for large obscuring columns. This is in
agreement with Georgantopoulos \& Georgakakis (2005) who claim that
this population comprises a significant fraction of sources where the
AGN emission is diluted by the host galaxy. We nevertheless find a
number of X-ray sources with early-type best-fit SED (type$<0.2$) and
high obscuring column densities ($N_H > \rm 10^{22} \, cm^{-2}$; see
Figure \ref{fig_nh_stype}). Although optical spectroscopy is not
available for these  systems they are candidates for the class of
obscured XBONGs similar to the `Fiore P3' source (Fiore et
al. 2000). 

Within the spectroscopic subsample there are a number of low-$z$
sources with unabsorbed X-ray spectra, broad emission-lines and
optical colours somewhat redder than those of typical QSOs. This
suggests either dust reddening and/or optical continuum dominated by
the host galaxy. For 7 of these sources flux calibrated optical
spectra are available allowing estimates of the optical reddening of
the broad-line region using the $\rm H\alpha / H\beta$ emission line
ratio. Table \ref{tab2} lists the  resulting $E(B-V)$ along with the
corresponding $N_H$ assuming a Galactic dust--to--gas ratio (Bohlin,
Savage \& Drake 1978). For some sources the column density listed in
this table is higher than the upper limit estimated from the X-ray
spectrum. This may suggest either anomalous dust--to--gas ratios
(Maiolino et al. 2001; Maiolino, Marconi \& Oliva 2001;
Georgantopoulos et al. 2003) or the  presence of a warm-absorber
(Crenshaw, Kraemer \& George 2003). We note however, that stellar
absorption from the host galaxy may affect the measured Balmer
emission line fluxes, particularly the $H\beta$, overestimating the
optical reddening and hence, the corresponding $N_H$. Nevertheless,
only  2 of the 7 broad-line sources show evidence for large
reddening. This suggests that dust extinction alone cannot fully
explain the red colours of all the spectroscopically identified broad
line systems.     

We explore the possibility of stellar light contamination of
the optical colours using the prescription below: for the intrinsic
spectrum of the central engine we adopt the SDSS composite QSO SED
(Vanden Berk et al. 2001).  We then estimate the contribution of the
AGN to the broad band magnitudes under the conservative assumption
that the observed $u$-band flux is entirely due to AGN emission. We
shift the SDSS composite QSO SED by the appropriate redshift and
redden it by the observed  extinction listed in Table \ref{tab2},
adopting the Galactic extinction law. This spectrum is then scaled so
that the integrated $u$-band flux matches the observed one. Finally,
for the scaled spectrum we estimate the flux through the SDSS $r$-band
and compare it with the observed one. The estimated fraction of QSO
emission in the $r$-band as well as the corresponding
$\alpha_{OX}$\footnote{we define $\alpha_{OX}= \log (f_{\rm
2500}/f_{\rm 2\,keV})/2.605$, where $f_{\rm 2500}$ and $f_{\rm 
2\,keV}$ is the flux density at $2500\AA$ and 2\,keV respectively.}
indices are listed in Table \ref{tab2}. We note that both the
resulting flux ratios and the $\alpha_{OX}$ are conservative upper
limits since we assume that all the observed $u$-band flux originates
from the central engine. With the exception of the two most obscured
sources, the AGN upper-limit contribution to the $r$-band is in the
range $40-60$ per cent. For the two sources with the highest $E(B-V)$
estimate in Table \ref{tab2} the estimated $\alpha_{OX}$ upper limits
are  unrealistically high and do not place any constraints. Adopting a
more realistic $\alpha_{OX}$ index of 1.5 (Brandt, Laor \& Wills 2000)
for these two systems we estimate an AGN contribution of $\la 1$ per
cent in the $r$-band. This increases to an upper limit of $\approx 10$
per cent in the extreme case of $\alpha_{OX}=2.2$, i.e. typical of
obscured systems (e.g. BAL QSOs; Brandt et al. 2000).

\begin{figure}
\centerline{\psfig{figure=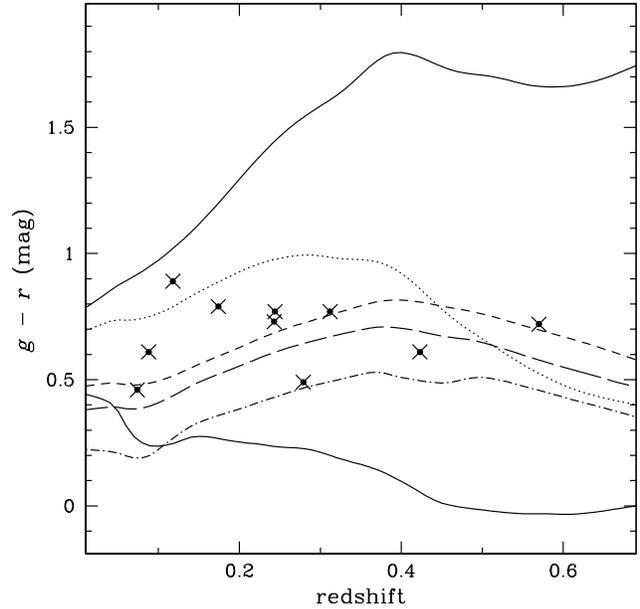,width=3.5in,angle=0}}
\caption
 {$g-r$ colour as a function of redshift for spectroscopically
 identified broad-line systems in our sample. The curves are the
 colour-$z$ tracks of galaxy/QSO composite spectra with fractional
 contribution of the galaxy emission at $\rm 6000\AA$ fixed to 80 per
 cent. Different galaxy types are plotted: the dotted line is for
 E/S0, the short-dashed line corresponds to Sbc, the long dashed and 
 dot-dashed lines are for Scd and Im types respectively. The
 upper and lower continuous lines represent pure E/S0 and QSO
 spectra and are plotted for comparison. 
 }\label{fig_composite}
\end{figure}

Figure \ref{fig_composite} shows how the $g-r$ colours of the SDSS QSO
spectrum are modified at different redshifts by superimposing galaxy
SEDs spanning the range E/S0--Im at a fixed galaxy/AGN $r$-band light
ratio of 80 per cent. Templates with a significant old stellar
component (e.g. E/S0, Sbc) can reproduce the observed  $g-r$ colours
of most sources. Late-type galaxy templates (e.g. Scd, Im) produce
composite spectra that are consistent with the colours of some sources
in Figure \ref{fig_composite}. These late-type SEDs however, require
on average larger galaxy/AGN light fractions ($>80$ per cent) to
reproduce the mean $g-r$ colours of the sources in Figure
\ref{fig_composite}. We conclude that dilution of the optical AGN
emission by stellar light is a possible explanation for the observed
red colours of moderate-$z$ broad-line systems if the galaxy/AGN ratio
is larger than $\approx 80$ per cent in the $r$-band.

\begin{table*}
\footnotesize
\begin{center}
\begin{tabular}{l cc cc c}
ID    & $\rm H\alpha/H\beta$ & $E(B-V)$ &  $N_H$$^{1}$           & $\alpha_{OX}$ & $r$-band \\
      &                      &          & ($\rm cm^{-2}$)  &               & AGN fraction \\
\hline
 61     &  2.9                & 0.00    & $-$              &  $<$1.72      & $<$0.54 \\
 66     &  3.6                & 0.13    & $6.5\times10^{20}$ &  $<$1.52    & $<$0.46\\
 06     &  2.7                & 0.00    & $-$              &  $<$1.14      & $<$0.40\\
 08     &  4.7                & 0.34    & $1.7\times10^{21}$ &  $<$1.67    & $<$0.56\\
 45     &  18.2               & 1.47    & $7.4\times10^{21}$ &  $<$2.77    & $<$3.36$^\star$\\
 51     &  20.9               & 1.58    & $7.9\times10^{21}$ &  $<$2.98    & $<$12.01$^\star$\\
 83     &  2.8                & 0.00    & $-$              &  $<$1.40      & $<$0.60\\
\hline
\multicolumn{6}{l}{$^1$The $N_H$ listed here is estimated from
	$E(B-V)$ adopting the Galactic}\\ 
\multicolumn{6}{l}{ dust--to--gas ratio (Bohlin, Savage \& Drake 1978)}\\   
\multicolumn{6}{l}{$^\star$For these two systems the AGN emission in the
$r$-band is estimated.}\\
\multicolumn{6}{l}{to be higher than the observed flux.} \\
\end{tabular}
\end{center}
\caption{Optical and X-ray properties of the 7 broad emission-line
sources with flux calibrated spectra}\label{tab2} 
\end{table*}

\subsection{Mean X-ray spectral properties}
\begin{figure} 
\centerline{\psfig{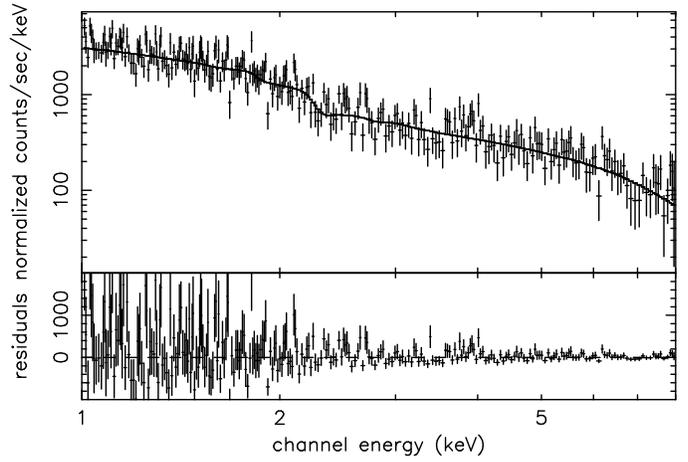}} 
\caption
{The EPIC-PN mean X-ray spectrum of the optically red hard X-ray selected
 sample. Only sources observed through the THIN {\it XMM-Newton}
 filter are presented here.  The continuous line is the single
 absorbed power-law (wabs*pow) model fit to the
 data.}\label{fig_xspec}      
\end{figure}

We attempt to constrain the co-added spectral properties of the red
optically extended sources to explore whether they are consistent with
those of the XRB ($\Gamma\approx1.4$; Lumb et
al. 2002; Kushino et al. 2002). The individual spectra of the
sample are merged using  the {\sc mathpha} task of {\sc ftools} to
produce 3 independent coadded spectral files for the PN, MOS1 and MOS2
detectors  respectively. Sources observed through the MEDIUM and THIN
filters of the {\it XMM-Newton} are merged separately. The combined
spectra are grouped to a minimum of 15  counts per bin to ensure that
Gaussian statistics apply. The auxiliary  files of individual sources
were combined using the {\sc addarf} task of {\sc ftools}. Using the
{\sc xspec} v11.2 software, we fit a single power-law to the data
absorbed by the Galactic column of $\rm 2\times 10^{20}\,cm^{-2}$
(wabs*pow). The fit is performed in the 1-8\,keV range. This is to
avoid energies below 1\,keV  where the XRB spectrum is likely to
change shape (McCammon \& Sanders 1990; Gendreau 
et al. 1995). The results for both the THIN (see also  
Figure \ref{fig_xspec}) and the MEDIUM filters respectively are
presented in Table \ref{tbl_res1}. The best-fit power-law index of the
two separate set of spectra are in fair agreement within the 90 per
cent confidence level. The mean X-ray spectrum of the population is
consistent with $\Gamma \approx 1.4$ similar to that of the diffuse
XRB underlying the importance of this population.  

\begin{table} 
\footnotesize
\begin{center} 
\begin{tabular}{cc cc c} 
\hline                        

{\it XMM-Newton} & number of & $\Gamma$ & $\chi^2_\nu$ & d.o.f. \\
filter           &  sources  &          &              &        \\
\hline
THIN     & 56  & $1.47^{+0.04}_{-0.04}$  & 1.08 & 562 \\
MEDIUM   & 27  & $1.40^{+0.05}_{-0.05}$  & 1.06 & 335 \\
\hline
\end{tabular}
\end{center}
\caption{
The mean X-ray spectral properties of the red optically extended
hard X-ray selected sample. Sources observed through the THIN and
MEDIUM {\it XMM-Newton} filters are merged separately. The errors
correspond to 90 per cent confidence level.
}\label{tbl_res1}
\normalsize
\end{table}     

\subsection{X-ray luminosity function}

In this section we derive the binned X-ray luminosity function of the
$z<0.4$ subsample using the  standard non-parametric $1/Vmax$ method
(Schmidt 1968). The choice of redshift cutoff is firstly to explore the
low-$z$ luminosity function of these systems and secondly
to minimise the incompleteness of our sample due to optically
unidentified sources, many of which are expected to lie at
$z\ga1$. For a given redshift and  X-ray luminosity interval the
binned luminosity function is estimated from the relation:  

\begin{equation}  
\phi \, d\log L = \sum_{i} \frac{1}{V_{max,i} (L,z)},
\end{equation} 

\noindent 
where $V_{max,i}(L,z)$ is the maximum survey volume available to the
source $i$ of the sample. This is estimated from the relation 

\begin{equation}  
V_{max,i}(L,z) = \frac{c}{Ho} \int_{z1}^{z2} \, \Omega(L,z)\, \frac{dV}{dz}\,dz\, dL,
\end{equation} 

\noindent 
where $dV/dz$ is the volume element per redshift interval $dz$. The
integration limits $z1$, $z2$ are the minimum and maximum redshifts
possible for a source of luminosity $L_X$ to remain within the flux
limits of the survey and to lie within the redshift bin. $\Omega(L,z)$
is the solid angle of the X-ray survey available to a source with
luminosity $L$ at a redshift $z$ (corresponding to a given flux in the
X-ray area curve). The logarithmic bin size of the luminosity
function varies so that each bin comprises approximately equal number
of sources, $N\approx10$.  The uncertainty of a given luminosity bin is
estimated from the relation:

\begin{equation}  
\delta \phi^2 = \sum_{i}     \left ( \frac{1}{V_{max,i} (L,z)} \right )^2.
\end{equation} 

\noindent
In addition to the non-parametric method above we also derive the
luminosity function using the parametric maximum likelihood method
(Tammann, Sandage \& Yahil 1979) assuming a conventional double power
law (e.g. Barger et al. 2005; Sazonov \&
Revnivtsev 2004) of the form:
 
\begin{equation}  
\frac{d\phi (L_X,z)}{d\log L_X}= 
\frac{ \phi_{\star} } { (L/L_{\star})^{g1}+ (L/L_{\star})^{g2} }, 
\end{equation} 

\noindent
where the characteristic luminosity evolves with redshift according to
the relation $L_{\star}=L_{o} \, (1+z)^A$. Our sample only covers
a narrow redshift range that does not allow constraining of the
evolution parameter $A$. We therefore, choose to fix $A=3$,
i.e. similar to that estimated by Barger et al. (2005) and Ueda et
al. (2003). We also assume that $\phi_{\star}$, the normalisation of
the luminosity function, is constant with redshift (e.g. no density
evolution; Barger et al. 2005). This is then estimated from the
relation

 \begin{equation}
 \phi_\star= N_{gal}/ \int \int \Omega(L,z) \, \Phi(L)/\phi_\star \,
 dL \, dV/dz \, dz  
 \end{equation} 

\noindent
where $N_{gal}$ is the total number of galaxies in the survey and
$\Omega(L,z)$ is the solid angle of the X-ray survey available to a
source with luminosity $L$ at a redshift $z$, i.e. the area curve at
different flux limits. The best-fit parameters are shown in Table
\ref{lftab}. The uncertainties correspond to the 68 per cent
confidence level and are estimated by fixing one parameter at a time
and then finding the maximum likelihood fit for the remaining
two. This is repeated by varying the value of the fixed parameter
between successive runs. The $1\sigma$ errors for that parameter
correspond  to $\delta L=2.7/2$ regions around the maximum likelihood 
fit. We note that the 68 per cent upper limit of the power-law index
$g_2$ is unconstrained. This is because our sample does not comprise
large number of luminous systems ($L_X>10^{44}\rm erg \, s^{-1}$) to
provide tight upper limits on $g_2$. Larger area coverage is required
to provide large numbers of such systems at $z<0.4$.  We also note
that heavily obscured AGNs ($N_H \ga  \rm 10^{23} \, cm^{-2}$) are
likely to be underrepresented in our 2-8\,keV selected sample.

\begin{figure}
 \centerline{\psfig{figure=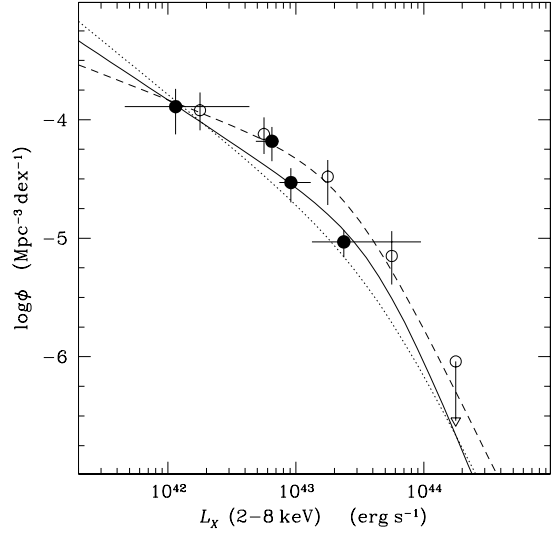,width=3in,angle=0}}
\caption
{ 
 The luminosity function (filled circles plotted at the median $L_X$
 of each bin) for the optically red X-ray 
 population with redshifts $z<0.4$. The continuous line is the maximum
 likelihood fit to the data.  This is compared to the Barger et
 al. (2005) X-ray luminosity function for all their sources in the range
 $0.1-0.4$ (open circles). The dashed line is the
 maximum likelihood fit to the Barger et
 al. (2005) data. The dotted line is the local RXTE luminosity function
 estimated by Sazonov \& Revnivtsev (2004). This is shifted to the
 2-8\,keV band assuming $\Gamma=1.8$ and evolved to $z=0.235$, the
 median redshift of our subsample, assuming luminosity evolution of
 the form $\propto (1+z)^{3}$.
 }
\label{lf2}
\end{figure}

The results of our analysis are shown in Figure \ref{lf2} along
with the maximum likelihood fit to the data.  Also shown in this plot
are the X-ray luminosity functions of Barger et al. (2005) for their
$0.1-0.4$ redshift bin and the local X-ray luminosity function of
Sazonov \& Revnivtsev (2004) shifted to the 2-8\,keV band assuming
$\Gamma=1.8$ and evolved to $z=0.235$, the median redshift of our
subsample, assuming luminosity evolution of the form $\propto
(1+z)^{3}$. There is  fair agreement between our luminosity function
and those derived in previous studies. This suggests that the red 
subsample comprises a sizable fraction of the  X-ray population at
$z<0.4$. In section 3.1 we indeed find that about 75 per cent of the
X-ray sources with optically extended counterparts (most likely to lie
at low-$z$) have red optical colours. This does not suggest that
broad-line AGNs, which typically have blue optical colours, are an
unimportant component of the X-ray population. Indeed, the colour
selected subsample studied here also comprises unboscured broad-line
AGNs, which appear optically redder most likely because of stellar
light contributing to the broad-band colours. The evidence above
underlines the importance of the red sub-population for 
understanding the different types of sources that make up the diffuse
X-ray background.  

Using the best-fit parametric form for the luminosity function we
derive the X-ray emissivity (luminosity per $\rm Mpc^3$)    
 
 \begin{equation}
 j_x=\int \Phi(L)~L~dL,
 \end{equation}

\noindent  
also taking into account the column density distribution of our
sources. We also estimate  the fractional contribution to the 2-8\,keV
X-ray background. The integrated galaxy X-ray flux is given by  
 
\begin{equation}\label{eq_xrb}
I=\frac{c}{4\pi H_{0}}\,
\int_{z1}^{z2}\frac{j_{x}~(1+z)^{p-\alpha_x}}{(1+z)(\Omega_m~(1+z)^3+\Omega_\Lambda)^{1/2}}\,   
 \, dz,  
\end{equation}
 
\noindent
where $j_x$ is the X-ray emissivity at $z=0$, $p$ is the adopted
evolution and $a_x=\Gamma - 1$, with $\Gamma=1.4$. We integrate
luminosities above $\rm 10^{41} erg\,s^{-1}$ assuming a pure
luminosity evolution with  $p=3$.  For the intensity of the XRB 
we adopt the normalisation of Vecchi et al. (1999). Integrating
equation \ref{eq_xrb} to $z=0.4$ and $z=1$ we find that the present
sample contributes about 6 and 17 per cent respectively of the
XRB. These fractions are somewhat uncertain depending on many
assumptions regarding the evolution of the different X-ray populations
and the absolute normalisation of the XRB intensity. We also note that the
sample used in this calculation comprises both type-I and II
Seyferts at moderate-$z$. Therefore, the XRB contributions estimated
above include many of the nearby counterparts of the high-$z$ type-I  
QSOs, which are responsible for a significant part of the XRB.

\begin{table*}
\footnotesize 
\begin{center} 
\begin{tabular}{ccc cc}
 $\log L_\star$ & $g_1$ & $g_2$$^\star$ & $\phi_\star$        & $j_x $ \\  
 ($\rm erg s^{-1}$) &       &       & ($\rm Mpc^{-3}\,dex^{-1}$)&
 ($\rm erg\,s^{-1}\,Mpc^{-3}$) \\
\hline 
 $43.37^{+0.99}_{-1.81}$ & $0.72_{-0.64}^{+0.24}$ &
$2.66_{-1.52}^{+\infty}$  & $1.03\times10^{-5}$ & $1.9\times 10^{38}$ \\
\hline 
\multicolumn{5}{l}{$^\star$The upper limit of the $g_2$ index is
 unconstrained. See text for details.} \\
\end{tabular}
\end{center}
\caption{Maximum likelihood parameters for the double-power law
luminosity function of red X-ray sources with $z<0.4$.}\label{lftab}
\end{table*}

\section{Discussion and conclusions}\label{discussion}

We combine data from the {\it XMM-Newton}/2dF survey with public {\it 
XMM-Newton} fields overlapping with the SDSS to explore the nature of 
the hard (2-8\,keV)  X-ray selected population hosted by red optically
extended sources. The optical properties above
suggest moderate redshifts ($z<1$) and either dust reddening or
optical continuum dominated by stellar emission. The significance of
this X-ray population is threefold: Firstly, these sources comprise a
significant fraction (about 75 per cent) of the  moderate-$z$ X-ray
population to the flux limit of our survey, $f_X ( \rm 2 - 8 \, keV ) = 2
\times 10^{-14} \rm \, erg \, s^{-1} \, cm^{-2}$. The X-ray luminosity
function of the subsample with $z<0.4$ is also found to be in fair
agreement with previous estimates at low-$z$, further suggesting that
our red colour selection picks the dominant X-ray population at these
redshifts. Secondly, using this luminosity function under the
assumption of luminosity evolution of the form $\propto (1+z)^3$
we estimate that these systems could be responsible for a
non-negligible component of the XRB to $z=1$, about 17 per
cent. Thirdly,  we argue that our sample comprises a large fraction of
the sources responsible for the spectral properties of the XRB. The
stacked X-ray spectrum of the red X-ray sources is found to be
consistent with a power-law with $\Gamma\approx1.4$, i.e. similar to
that of the XRB.  

Our sample comprises a mixed bag of objects including both absorbed
and unabsorbed systems. Limited spectroscopic information available
for our sample suggests that the X-ray absorbed population is mostly
associated with narrow-emission line galaxies (e.g. Seyfert-2
activity) while the unabsorbed sources show either broad
emission-lines (e.g. Seyfert-1s) or absorption lines (e.g. XBONGs). We
argue that the red colours of the majority of the sources in the
sample can be explained by the host galaxy stellar emission
significantly contributing to the optical continuum. 

For the obscured X-ray population previous studies indeed suggest that
the optical colours of this class of sources are dominated by the host
galaxy (Barger et al. 2002, 2003; Georgakakis et al. 2004a) with the AGN
emission most likely blocked from view by the obscuring screen of gas
and dust. In the case of  the unobscured sources in our sample 
with $L_X\approx 10^{42} - 10^{44} \rm \, erg \,  s^{-1}$ it is not
clear why their colours are redder that those of AGNs. For these
systems we estimate a {\it lower} limit for the stellar contribution
to the $r$-band optical continuum of $40-60$ per cent. We show
that an old (red) stellar population contributing up to 80 per cent in
the $r$-band can indeed make these systems appear redder than typical
AGNs. Higher fractions are however, required for young stellar 
populations. Severgnini et al. (2003) argue that the continuum
emission of an $L_X(2-8)=10^{43}\rm \, erg \, s^{-1}$ AGN can be
swamped by a galaxy brighter than $M_R=-22$\,mag. The median absolute
magnitude of our sample is $M_r=-22.32$\,mag, providing further 
support to the scenario of AGN dilution by stellar light. This
luminosity is about 1\,mag brighter than the characteristic absolute
magnitude of the luminosity function, $M_R^*=21.2$ (Blanton et
al. 2003; $\rm Ho=70 \, km \, s^{-1} \, Mpc^{-1}$). Higher optical
luminosities may be hard to understand in terms of stellar light
alone. In these most extreme systems (in term of absolute magnitude)
the AGN may dominate the optically luminosity and moderate dust
obscuration (consistent with the estimated low $N_H$ of the broad-line
systems) may be responsible  for the red optical colours. For example
in the case of Galactic dust-to-gas ratio a column density of $N_H
\approx  10^{21} \, \rm cm^{-2}$ corresponds to $g-r$ reddening
$E(g-r)\approx0.3$\,mag. 

Francis, Nelson \& Cutri (2004) used the  Two Micron All Sky Survey
(2MASS) to explore the fraction of red  AGNs at low-$z$ likely to
remain unidentified in UV/optical selected samples. They find that a
small fraction of the $J-K>1.2$\,mag selected population (about 1 per
cent) is associated with broad emission-line AGNs. They also argue
that host galaxy light contamination is responsible for the red
colours of at least half of these systems.   

Cutri et al. (2001) also identified broad line AGNs (including types
1, 1.5 and 1.8) in the 2MASS catalogue albeit with extremely red
colours ($J-K>2$\,mag), the vast majority of which lie at
$z\approx0.25$. Wilkes et al. (2002) explored the X-ray properties of
these  broad line AGN and found evidence for obscuration in the range
$N_H \approx 10^{21} - 10^{23} \rm \, cm^{-2}$. They argue that dust
reddening is most likely responsible for the red colours of these
sources. However, the Wilkes et al. (2002) AGNs are, on average,
strikingly underluminous at X-ray wavelengths relative to their
near-IR flux. This is unlike the broad-line  red sources studied here
that have $\log f_X /f_{opt}$ typical of  QSOs (see Figure
\ref{fig_fxfo}). Moreover, the median $N_H$ of the red X-ray sample
studied here, including a substantial number of  narrow-line systems
with high obscurations, is $\approx 10^{21} \rm \, cm^{-2}$. This is
lower than $\approx 10^{22} \rm \, cm^{-2}$ for the Wilkes et
al. (2002) 2MASS AGNs, although the majority of their systems have 
broad emission lines (types 1.8 or earlier) with only a small fraction
showing narrow emission lines. Focusing on the spectroscopically
identified broad-line sources in the present sample we find X-ray
spectral  properties suggesting little photoelectric absorption above
the Galactic value. We conclude that at least the spectroscopically 
identified red broad-line AGNs in our sample are not the same
population as those with very red optical/near-IR colours studied by
Wilkes et al. (2002) but similar to the less extreme (in terms of
$J-K$ colour) red AGNs discussed by Francis et al. (2004). We cannot
nevertheless, exclude the possibility that extreme objects like those
studied by Wilkes et al. (2002) are present in our
non-spectroscopically identified sample.

\section{Acknowledgments}
 We thank the anonymous referee for valuable comments and
 suggestions. AG and AA acknowledge funding by the European Union and
 the Greek Government  in the framework of the programme ``Promotion
 of Excellence in Technological Development and Research'', project 
 ``X-ray Astrophysics with ESA's mission XMM''.   

 Funding for the creation and distribution of the SDSS Archive has
 been provided by the Alfred P. Sloan Foundation, the Participating
 Institutions, the National Aeronautics and Space Administration, the
 National Science Foundation, the U.S. Department of Energy, the
 Japanese Monbukagakusho, and the Max Planck Society. The SDSS Web
 site is http://www.sdss.org/. The SDSS is managed by the
 Astrophysical Research Consortium (ARC) for the Participating
 Institutions. The Participating Institutions are The University of
 Chicago, Fermilab, the Institute for Advanced Study, the Japan
 Participation Group, The Johns Hopkins University, Los Alamos
 National Laboratory, the Max-Planck-Institute for Astronomy (MPIA),
 the Max-Planck-Institute for Astrophysics (MPA), New Mexico State
 University, University of Pittsburgh, Princeton University, the
 United States Naval Observatory, and the University of Washington.


\begin{thebibliography}{} 


{\bibitem{3} Barger A. J., Cowie L. L., Mushotzky R. F., Yang Y., Wang
W.-H., Steffen A. T., Capak P.,  2005, AJ, 129..578}  

{\bibitem{4} Barger, A. J. et al. 2003, AJ, 126, 632}


{\bibitem{5}Barger A. J., Cowie L. L., Brandt W. N., Capak P.,
Garmire G. P., Hornschemeier A. E., Steffen A. T., Wehner E. H.,
2002, AJ, 124, 1839}

{\bibitem{6} Bohlin R. C., Savage B. D., Drake J. F., 1978, ApJ, 224, 132}


{\bibitem{7}  Brandt W. N., et al., 2001, AJ, 122, 2810}



{\bibitem{9}  Brandt W. N., Laor A., Wills B. J., 2000, ApJ, 528, 637}

{\bibitem{10}  Carrera F. J., Barcons X., Butcher J. A., Fabian A. C.,
Lahav O., Stewart G. C., Warwick R. S.,  1995, MNRAS, 275, 22}    

{\bibitem{10a} Cash W., 1979, ApJ, 228, 939}

{\bibitem{11} Coleman G. D., Wu C.-C., Weedman D. W., 1980, ApJ, 43,
393}

{\bibitem{12} Comastri A., et al., 2002,  ApJ, 571, 771}


{\bibitem{14} Crenshaw D. M., Kraemer S. B., George I. M., 2003, ARA\&A,
41, 117}






{\bibitem{19} Cristiani S., et al., 2004, ApJ, 600, L119}


{\bibitem{19a} Csabai I., et al., 2003, ApJ, 125, 580}

{\bibitem{20} Cutri R. M., Nelson B. O., Kirkpatrick J. D., Huchra J. P., Smith
P. S., 2001, AAS, 198, 3317} 

{\bibitem{21} Downes A. J. B., Peacock J. A., Savage A., Carrie D. R.,
1986, MNRAS, 218, 31}

{\bibitem{22} Fiore F., et al.,  2003, A\&A, 409, 79}

{\bibitem{23} Fiore F., et al., 2000, New Astronomy, 5, 143}

{\bibitem{24}Francis P. J., Nelson B. O., Cutri R. M., 2004, AJ, 127, 646}



{\bibitem{25a} Gendreau et al., 1995, PASJ, 47, 5L}


{\bibitem{26} Georgakakis A., Hopkins A. M., Afonso J., Sullivan
M., Mobasher B., Cram L. E., 2004a, MNRAS, 354, 127}

{\bibitem{26a} Georgakakis A., et al., 2004b, MNRAS, 349, 135}

{\bibitem{27} Georgakakis A., Georgantopoulos I., Stewart G. C.,
Shanks T., Boyle B. J., 2003, MNRAS, 344, 161}

{\bibitem{28} Georgantopoulos I., Georgakakis A., 2005, MNRAS, in press,
astro-ph/0412335}

{\bibitem{29} Georgantopoulos I., Georgakakis A., Akylas A., Stewart
G. C., Giannakis O., Shanks T., Kitsionas S., 2004, MNRAS, 352, 91}

{\bibitem{30} Georgantopoulos I., Zezas A., Ward M. J., 2003, ApJ, 584,
129} 



{\bibitem{30a} Georgantopoulos I., Georgakakis A., Stewart G. C.,
  Akylas A., Boyle B. J., Shanks T., Griffiths R. E., 2003, MNRAS,
  342, 321}  

{\bibitem{31} Giacconi R., et al.,  2002, ApJS, 139, 369}


{\bibitem{33}  Grogin N. A., et al., 2003, ApJ, 595, 685}



{\bibitem{35} Hatziminaoglou E., Mathez G., Pell\'o R., 2000, A\&A, 359,
9}

{\bibitem{36} Jahoda K., Mushotzky R. F., Boldt E., Lahav O.,
1991, ApJ, 378, 37L} 

{\bibitem{37} Kitsionas S., Hatziminaoglou S., Georgakakis A.,
Georgantopoulos I., 2005, A\&A, in press, astro-ph/0406346}

{\bibitem{38}Koekemoer A. M., et al., 2002, ApJ, 567, 657} 


{\bibitem{40} Kushino A., Ishisaki Y., Morita U., Yamasaki N. Y., Ishida
M., Ohashi T., Ueda Y., 2002, PASJ, 54, 372}

{\bibitem{41} Lahav O., et al., 1993,  Natur, 364, 693}


{\bibitem{41a}Lehmann I., et al., 2001, A\&A, 371, 833}
 
{\bibitem{42} Lumb D. H., Warwick R. S., Page M., De Luca A., 2002,
A\&A, 389, 93}

{\bibitem{43} Maiolino R., Marconi A., Oliva E., 2001, A\&A, 365, 37} 

{\bibitem{44} Maiolino R., Marconi A., Salvati M., Risaliti G.,
Severgnini P., Oliva E., La Franca F., Vanzi L.,  2001, A\&A, 365, 28} 



{\bibitem{45a} McCammon D., Sanders W. T.,  1990, ARA\&A, 28, 657}


{\bibitem{47} Miyaji T., Lahav O., Jahoda K., Boldt E., 1994,
ApJ, 434, 424}


{\bibitem{48} Mobasher B., et al., 2004, ApJ, 600, 167L}

{\bibitem{49} Moran E. C., Lehnert M. D., Helfand D. J., 1999, ApJ, 526, 649}

{\bibitem{50} Moran E., Filippenko A. V., Chornock R.,  2002, ApJ,
579L, 71}





{\bibitem{54} Reynolds C. S., 1997, MNRAS, 286, 513}

{\bibitem{55} Richards G. T., et al., 2002, AJ, 123, 2945}

{\bibitem{56} Rosati P., et al., 2002, ApJ, 566, 667}


{\bibitem{58} Sazonov S. Y., Revnivtsev M. G., 2004, A\&A, 423, 469}

{\bibitem{59} Severgnini P., et al., 2003, A\&A, 406, 483}

{\bibitem{60} Schmidt, M. 1968, ApJ, 151, 393}



{\bibitem{62} Stocke J. T., Morris S. L., Gioia I. M., Maccacaro T.,
Schild R., Wolter A., Fleming T. A., Henry J. P., 1991, ApJS, 76, 813}


{\bibitem{63} Stoughton C., et al.,  2002, AJ, 123, 485.}

\bibitem{64} Tammann, G.A., Yahil, A., Sandage, A., 1979, ApJ, 234, 775  

{\bibitem{65} Treister E., et al.,  2004, ApJ, 616, 123}

{\bibitem{66} Ueda Y., Akiyama M., Ohta K., Miyaji T., 2003, ApJ, 598,
886} 

{\bibitem{67} Vanden Berk D. E., et al.,  2001, AJ, 122, 549}


{\bibitem{67a} Vecchi A., Molendi S., Guainazzi M., Fiore F., Parmar
A. N., 1999, A\&A, 349, 73L}


{\bibitem{69} White R. L., Helfand D. J., Becker R. H., Gregg M. D.,
Postman M., Lauer T. R., Oegerle W., 2003, AJ, 126, 706}

{\bibitem{70} Wilkes B. J., Schmidt G. D., Cutri R. M., Ghosh H., 
Hines D. C., Nelson B., Smith P. S.,  2002, ApJ, 564, 65L}

{\bibitem{71} York D. G., et al., 2000, AJ, 120, 1579.}


{\bibitem{72}  Zezas A. L., Georgantopoulos I., Ward M. J.,  1998,
MNRAS, 301, 915}

\end{thebibliography}
\end{document}